\documentclass[12pt]{article}
\usepackage{epsfig}
\usepackage{amssymb}
\usepackage{amsmath}
\usepackage{amsfonts}

\newcommand{\R}{\mathbb{R}}
\newcommand{\C}{\mathbb{C}}
\newcommand{\Z}{\mathbb{Z}}

\newcommand{\fp}{\mathfrak{p}}

\newcommand{\fw}{\mathfrak{w}}

\newcommand{\fz}{\mathfrak{z}}

\newcommand{\fP}{\mathfrak{P}}

\newcommand{\cJ}{\mathcal{J}}

\newcommand{\be}{\begin{equation}}
\newcommand{\ee}{\end{equation}}
\newcommand{\bea}{\begin{eqnarray}}
\newcommand{\eea}{\end{eqnarray}}
\newcommand{\nn}{\nonumber}

\newcommand{\ed}{\end{document}}

\newcommand{\lpb}{\{\!\!\{}
\newcommand{\rpb}{\}\!\!\}}

\newcommand{\rx}{{\rm x}}

\newcommand{\rp}{{\rm p}}

\newcommand{\np}{\newpage}

\begin{document}

\title{Real Description of Classical Hamiltonian Dynamics
Generated by a Complex Potential}

\author{\\
Ali Mostafazadeh
\\
\\
Department of Mathematics, Ko\c{c} University,\\
34450 Sariyer, Istanbul, Turkey\\ amostafazadeh@ku.edu.tr}
\date{ }
\maketitle

\begin{abstract} Analytic continuation of the classical dynamics
generated by a standard Hamiltonian, $H=\frac{\rp^2}{2m}+v(\rx)$,
into the complex plane yields a particular complex classical
dynamical system. For an analytic potential $v$, we show that the
resulting complex system admits a description in terms of the
phase space $\R^4$ equipped with an unconventional symplectic
structure. This in turn allows for the construction of an
equivalent real description that is based on the conventional
symplectic structure on $\R^4$, and establishes the equivalence of
the complex extension of classical mechanics that is based on the
above-mentioned analytic continuation with the conventional
classical mechanics. The equivalent real Hamiltonian turns out to
be twice the real part of $H$, while the imaginary part of $H$
plays the role of an independent integral of motion ensuring the
integrability of the system. The equivalent real description
proposed here is the classical analog of the equivalent Hermitian
description of unitary quantum systems defined by  complex,
typically ${\cal PT}$-symmetric, potentials.

\vspace{5mm}



\end{abstract}





\section{Introduction}

The recent discovery that the standard quantum Hamiltonian
operators,
    \be
    \hat H=\frac{\hat{p}^2}{2m}+v(\hat x),
    \label{H-quantum}
    \ee
with certain complex potentials such as $v(x)=ix^3$ have a purely
real spectrum \cite{bander-98} has triggered a thorough
investigation of quantum systems defined by such Hamiltonians. An
important outcome of this investigation is that at least for the
cases that the spectrum is discrete the reality of the spectrum is
not only necessary \cite{p1} but also sufficient \cite{p2} for the
existence of a positive-definite inner product that renders the
quantum dynamics unitary \cite{p3}.\footnote{This inner product is
not unique \cite{p4,jmp-2003}. A particular example is the ${\cal
CPT}$-inner product proposed in \cite{bbj}.} As shown in
\cite{p2,p3} under these conditions the Hamiltonian turns out to
be quasi-Hermitian \cite{quasi}. This in turn leads to another
crucial finding namely that the resulting unitary quantum system
admits an equivalent Hermitian description \cite{jpa-2003}. The
latter can be used to define an underlying classical Hamiltonian
system whose pseudo-Hermitian quantization yields the initial
quantum system \cite{jpa-2004b,jpa-2005b,jpa-2006}. This
construction of an underlying classical system for the quantum
Hamiltonian (\ref{H-quantum}) is fundamentally different from a
direct association of this Hamiltonian operator with the complex
classical Hamiltonian
    \be
    H=\frac{\rp^2}{2m}+v(\rx).
    \label{H-class}
    \ee
The purpose of this article is to investigate the possibility of a
real description of the complex dynamical systems generated by the
Hamiltonians of the form (\ref{H-class}) where $v:\C\to\C$ is an
analytic function.

Specific examples of these complex classical systems have been
studied in
\cite{bender-99,nanayakkara,curtright-mezincescu,bender-chen-darg-milton},
and a complex phase space approach has been proposed in
\cite{kaushal-singh}. But, to the best of our knowledge, a
thorough investigation of the associated symplectic structure(s)
and the relation to the conventional real classical dynamical
systems has not been reported previously.

For a complex-valued potential the Hamiltonian dynamics defined by
(\ref{H-class}) takes place in a complex phase space. Hence we
shall use $\fz$ and $\fp$ to denote the complex dynamical
phase-space coordinates $\rx$ and $\rp$, respectively. In this
notation, the classical Hamiltonian reads
    \be
    H=\frac{\fp^2}{2m}+v(\fz),~~~~~~~\fz,\fp\in\C.
    \label{H}
    \ee
We will also introduce
    \bea
    &&x:=\Re(\fz),~~~~~~~~~~y:=\Im(\fz),
    ~~~~~~~~~~p:=\Re(\fp),~~~~~~~~~~q:=\Im(\fp),
    \label{id1}\\
    &&v_r(x,y):=\Re(v(x+iy)),~~~~~~~~~~~~~~~~~v_i(x,y):=\Im(v(x+iy)),
    \label{id2}\\
    &&H_r:=\Re(H)=\frac{p^2-q^2}{2m}+v_r(x,y),~~~~~
      H_i:=\Im(H)=\frac{pq}{m}+v_i(x,y),
    \label{id3}
    \eea
where $\Re$ and $\Im$ stand for the real and imaginary parts of
their argument, respectively. Note also that because $v$ is
assumed to be a (complex) analytic function, $v_r$ and $v_i$
satisfy the Cauchy-Riemann conditions:
    \be
    \partial_x v_r(x,y)=\partial_y v_i(x,y),~~~~
    \partial_y v_r(x,y)=-\partial_x v_i(x,y).
    \label{C-R}
    \ee

\section{Compatible Symplectic Structures}

The complex Hamiltonian~(\ref{H}) defines a dynamics in the
complex phase space $\C^2$ according to the Hamilton's equations
    \be
    \dot\fz=\partial_{\fp}
    H=\frac{\fp}{m},~~~~\dot\fp=
    -\partial_{\fz}H=-\partial_{\fz}v(\fz),
    \label{dyn-eq}
    \ee
where a dot denotes a time-derivative, and the time parameter $t$
is assumed to take real values. Our aim in this section is to
determine the symplectic structures \cite{arnold} on the phase
space $\fP=\C^2$ that are compatible with the dynamical
equations~(\ref{dyn-eq}). In other words, we wish to construct a
Poisson-like bracket (an antisymmetric, non-degenerate, bilinear
form also called skew inner product \cite{arnold})
$\lpb\cdot,\cdot\rpb$ in terms of which (\ref{dyn-eq}) takes the
form
    \be
    \dot\fz=\lpb\fz,H\rpb,~~~~~~\dot\fp=\lpb\fp,H\rpb.
    \label{bp}
    \ee
Before addressing this problem, however, we shall first show that
the choice of the standard symplectic structure, i.e., setting
$\fP=\R^4$ and endowing it with the standard symplectic structure,
is not consistent with the dynamical equations~(\ref{dyn-eq}).

The standard symplectic structure on $\R^4=\C^2$ is defined by the
conventional Poisson bracket $\{\cdot,\cdot\}$ according to
    \bea
    \{A,B\}&:=&
    (\partial_xA\:\partial_pB+\partial_yA\:\partial_qB)-
    (A\leftrightarrow B)
    \nn\\&=&
    2(\partial_{\fz}A\:\partial_{\fp^*}B+
    \partial_{\fz^*}A\:\partial_{\fp}B)-(A\leftrightarrow B),
    \label{conv-pb}
    \eea
where $A,B:\fP\to\C$ are smooth functions, $(A\leftrightarrow B)$
stands for the preceding terms with $A$ and $B$ exchanged, and we
have made use of the identities
    \be
    \partial_{\fz}=\frac{1}{2}(\partial_x-i\partial_y),~~~~~
    \partial_{\fz^*}=\frac{1}{2}(\partial_x+i\partial_y).
    \label{partial}
    \ee
Clearly, in view of (\ref{H}) and (\ref{conv-pb}),
    \[\dot\fz=\{\fz,H\}=0,~~~~~~~~~\dot\fp=\{\fp,H\}=0.\]
Hence, a symplectic structure that is consistent with the
dynamical equations~(\ref{dyn-eq}), if exists, is not the standard
one. It is this observation that motivates the search for finding
dynamically compatible nonstandard symplectic structures. To the
best of our knowledge the first step in this direction is taken in
\cite{curtright-mezincescu} where the authors briefly discuss the
issue and give a special class of compatible symplectic
structures. In the following we offer a thorough and systematic
investigation of the compatible symplectic structures.

To construct a compatible symplectic structure on the phase space
we recall using the defining properties of $\lpb\cdot,\cdot\rpb$
that
    \be
    \lpb A,B\rpb=\sum_{i,j=1}^4 \cJ_{ij}~
    \partial_{\fw_i}\!A\;\partial_{\fw_j}\!B,
    \label{syp-st-c}
    \ee
where $\fw_1:=\fz,\fw_2:=\fp,\fw_3:=\fz^*,\fw_4:=\fp^*$, and
$\cJ_{ij}$ are components of a symplectic form $\omega_{_\cJ}$ or
the entries of the associated invertible antisymmetric matrix
$\cJ$. The latter is sometimes called a symplectic matrix
\cite{marsden-ratiu}.

Imposing the reality condition,
    \be
    \lpb A,B\rpb^*=\lpb A^*,B^*\rpb,
    \label{reality}
    \ee
and requiring the compatibility with the dynamical
equations~(\ref{dyn-eq}) and non-degeneracy of $\omega_{_\cJ}$
(equivalently invertibility of $\cJ$), we find
    \bea
    &&\lpb \fz,\fp\rpb=1,~~~~
    \lpb \fz,\fz^*\rpb=ia,~~~~\lpb
    \fz,\fp^*\rpb=\alpha,
    \label{e1}\\
    &&\lpb \fp,\fz^*\rpb=-\alpha^*,~~~~\lpb \fp,\fp^*\rpb=ib,
    ~~~~\lpb \fz^*,\fp^*\rpb=1,
    \label{e2}
    \eea
where $a,b\in\R$ and $\alpha\in\C$ such that $|\alpha|^2-ab\neq
1$. Eqs.~(\ref{e1}) and (\ref{e2}) together with the antisymmetry
of $\lpb\cdot,\cdot\rpb$ determines the latter in terms of the
free parameters $a,b,\alpha$. Specifically, $\lpb\cdot,\cdot\rpb$
satisfies (\ref{syp-st-c}) with $\cJ$ given by
    \be
    \cJ=\left(\begin{array}{cccc}
    0 & 1 & ia & \alpha\\
    -1 & 0 & -\alpha^* & ib\\
    -ia & \alpha^* & 0 & 1\\
    -\alpha & -ib & -1 & 0\end{array}\right).
    \label{cj=}
    \ee

In order to see if $\lpb\cdot,\cdot\rpb$ defines a real symplectic
structure on $\R^4$, we introduce
    \be
    w_1:=x,~~~~~~w_2:=p,~~~~~~w_3:=y,~~~~~~w_4=q,
    \label{w=}
    \ee
and express $\lpb\cdot,\cdot\rpb$ as
    \be
    \lpb A,B\rpb=\sum_{i,j=1}^4 J_{ij}~
    \partial_{w_j}A\:\partial_{w_j}B,
    \label{syp-st-r}
    \ee
where $J_{ij}$ depend on $\cJ_{ij}$. A straightforward calculation
using (\ref{partial}), (\ref{syp-st-c}), (\ref{cj=}), and
(\ref{syp-st-r}) identifies $J_{ij}$ with the entries of
    \be
    J=\frac{1}{2}\left(\begin{array}{cccc}
    0 & 1+\alpha_r & -a & -\alpha_i\\
    -(1+\alpha_r) & 0 & -\alpha_i & -b\\
    a & \alpha_i & 0 & -1+\alpha_r\\
    \alpha_i & b & 1-\alpha_r & 0\end{array}\right),
    \label{rj=}
    \ee
where $\alpha_r:=\Re(\alpha)$ and  $\alpha_i:=\Im(\alpha)$. As
seen from (\ref{rj=}), $J$ is a real invertible antisymmetric
(symplectic) matrix, and $\lpb\cdot,\cdot\rpb$ defines a genuine
symplectic structure on $\R^4$ that is by construction compatible
with the dynamical equations (\ref{dyn-eq}). It is not difficult
to see that indeed (\ref{rj=}) is the most general symplectic
matrix with these properties.

The standard symplectic structure, that is defined using the
symplectic matrix
    \be
    J_{\rm st}=\left(\begin{array}{cccc}
    0 & 1 & 0 & 0 \\
    -1 & 0 & 0 & 0\\
    0 & 0 & 0 & 1\\
    0 & 0 & -1 & 0\end{array}\right),
    \label{rj=standard}
    \ee
does not fulfil (\ref{rj=}). The simplest example of the allowed
symplectic matrices (\ref{rj=}) is
    \be
    J_0=\frac{1}{2}\left(\begin{array}{cccc}
    0 & 1 & 0 & 0 \\
    -1 & 0 & 0 & 0\\
    0 & 0 & 0 & -1\\
    0 & 0 & 1 & 0\end{array}\right),
    \label{rj=zero}
    \ee
which corresponds to the choice $a=b=\alpha=0$.

\section{Equivalent Formulation Using Standard\\ Symplectic
Structure}

The fact that $J_{\rm st}$ fails to belong to the class of
symplectic matrices (\ref{rj=}) does not mean that the latter are
associated with fundamentally different theories. According to the
well-known uniqueness theorem for the symplectic structures on
$\R^{2n}$, every symplectic structure is isomorphic to the
standard one \cite{arnold}.

For the case at hand, it is not difficult to find a similarity
transformation $J\to J'=S^{-1} J S$, by a real orthogonal matrix
$S$, that maps $J$ to\footnote{$J$ has four linearly independent
eigenvectors ${\vec v}_+, {\vec v}_+^*,\vec v_-,{\vec v}_-^*$. The
columns of $S$ are unit vectors aligned along the real and
imaginary parts of ${\vec v}_\pm$. $S$ is orthogonal, because the
eigenvectors of $J$ are orthogonal.}
    \be
    J':=\left(\begin{array}{cccc}
    0 & r_+ & 0 & 0 \\
    -r_+ & 0 & 0 & 0\\
    0 & 0 & 0 & r_-\\
    0 & 0 & -r_- & 0\end{array}\right),
    \label{j-prime}
    \ee
where
    \[r_\pm:=\sqrt{\frac{1}{8}\left(a^2+b^2+2(|\alpha|^2+1)\pm
    \sqrt{[(a+b)^2+4][(a-b)^2+4|\alpha|^2]}\right)}\in\R^+.\]
Hence the following new coordinates in $\R^4$ serve as the
symplectic (Darboux) coordinates associated with the symplectic
matrix $J$.
    \be
    x_1=r_+^{-1/2}\sum_{k=1}^4 S_{k1}w_k,~~~
    p_1=r_+^{-1/2}\sum_{k=1}^4 S_{k2}w_k,~~~
    x_2=r_-^{-1/2}\sum_{k=1}^4 S_{k3}w_k,~~~
    p_2=r_-^{-1/2}\sum_{k=1}^4 S_{k4}w_k,
    \label{tilde}
    \ee
where we used the fact that $S$ is orthogonal.

As explicit expressions for the symplectic coordinates
(\ref{tilde}) are complicated, we will here suffice to present
them only for the simplest case, namely $a=b=\alpha=0$, that
corresponds to the symplectic matrix $J_0$. In this case, we have
    \be
    x_1=\sqrt{2}\, w_1=\sqrt 2\, x,~~~
    p_1=\sqrt{2}\, w_2=\sqrt 2\, p ,~~~
    x_2=\sqrt 2\, w_4= \sqrt 2\, q,~~~
    p_2=\sqrt 2\, w_3=\sqrt 2\, y,
    \label{tilde-zero}
    \ee
which are essentially identical with those initially considered in
\cite{xa}. See also \cite{kaushal-singh}.\footnote{Note that
unlike in \cite{xa,kaushal-singh} where these coordinates were
introduced essentially for convenience, we offer a systematic
derivation of them based on the uniqueness theorem for symplectic
structures.}

Having obtained a set of symplectic coordinates associated with a
dynamically compatible symplectic structure, we can express the
dynamical equations (\ref{dyn-eq}) in terms of a set of standard
Hamilton equations, namely
    \bea
    \dot x_1&=&\{x_1,h\}=\frac{p_1}{m},~~~~~~~~
    \dot x_2=\{x_2,h\}=2~\partial_{p_2}v_r(
    \mbox{\footnotesize $2$}^{-\frac{1}{2}}x_1,
    \mbox{\footnotesize $2$}^{-\frac{1}{2}}p_2),
    \label{dyn-1}\\
    \dot p_1&=&\{p_1,h\}=-2~ \partial_{x_1}v_r(
    \mbox{\footnotesize $2$}^{-\frac{1}{2}}x_1,
    \mbox{\footnotesize $2$}^{-\frac{1}{2}}p_2),~~~~~~~~
    \dot p_2=\{p_2,h\}=\frac{x_2}{m},
    \label{dyn-2}
    \eea
for the real Hamiltonian
    \be
    h:=\frac{p_1^2-x_2^2}{2m}+2\,v_r(
    \mbox{\footnotesize $2$}^{-\frac{1}{2}}x_1,
    \mbox{\footnotesize $2$}^{-\frac{1}{2}}p_2)=2H_r.
    \label{h}
    \ee
One can check using the (\ref{H})~--~(\ref{C-R}) and
(\ref{partial}) that (\ref{dyn-1})~--~(\ref{dyn-2}) are equivalent
to (\ref{dyn-eq}).\footnote{After the completion of this project
it was brought to our attention that the observation that the real
part of a complex analytic Hamiltonian can generate the dynamics
in the coordinates (\ref{tilde-zero}) was previously made in
\cite{xa}.} In particular, the structure of the trajectories in
the $x$-$y$ (equivalently $x_1$-$p_2$) plane for the ${\cal
PT}$-symmetric potentials $v(\fz)=-(i\fz)^n$ (with
$n\in\Z$)\footnote{For non-integer $n$ this potential is not an
entire function and special care needs be taken whenever a
trajectory crosses a branch cut.} and $v(\fz)=\sum_{k>0}\mu_k
e^{ik\fz}$ (with $\mu_k\in\R$) that are respectively examined in
\cite{bender-99,bender-chen-darg-milton} and
\cite{curtright-mezincescu} can be obtained using the real
Hamiltonian (\ref{h}).

As expected $H_r$ which is half the Hamiltonian $h$ is an integral
of motion. The same is true about
    \be
    H_i=\frac{x_2p_1}{2m}+v_i(
    \mbox{\footnotesize $2$}^{-\frac{1}{2}}x_1,
    \mbox{\footnotesize $2$}^{-\frac{1}{2}}p_2),
    \label{H-i=}
    \ee
i.e., $\dot{H}_i=\{H_i,h\}=0$.\footnote{It is a straightforward
exercise to show using (\ref{dyn-1}), (\ref{dyn-2}), and
(\ref{H-i=}) that $\dot{H}_i=0$.} It provides an independent
integral of motion for the system that ensures its integrability
via Liouville's theorem \cite{arnold}. What has been done in the
recent studies of ${\cal PT}$-symmetric potentials
\cite{bender-99,nanayakkara,bender-chen-darg-milton} is to set the
value of $H_i$ to zero and study the behavior of the solutions
satisfying this constraint. Table~1 gives the explicit form of the
real Hamiltonian $h$ and the invariant $H_i$ for some typical
${\cal PT}$-symmetric potentials.
    \begin{table}[h]
    \vspace{.5cm}
    \begin{center}
  \begin{tabular}{||c|c|c||}
  \hline \hline
  $v(\fz)$ & $h(x_1,x_2,p_1,p_2)$ & $H_i(x_1,x_2,p_1,p_2)$ \\
  \hline\hline
  $i\,\fz$ & $p_1^2-p_2/\sqrt 2-x_2^2$ & $x_2p_1+x_1/\sqrt 2$\\
  \hline
  $\fz^2$ & $p_1^2-p_2^2+x_1^2-x_2^2$ & $x_2p_1+x_1p_2$\\
  \hline
  $i\fz^3$ & $p_1^2+(p_2^3-3x_1^2p_2)/\sqrt 2-x_2^2$ &
  $x_2p_1+(x_1^3-3x_1p_2^2)/2\sqrt 2$\\
  \hline
  $-\fz^4$ & $p_1^2-(x_1^4-6x_1^2p_2^2+p_2^4)/2-x_2^2$ &
  $x_2p_1-x_1^3p_2-x_1p_2^3$\\
  \hline
  $e^{i\fz}$ & $p_1^2+2\:e^{-p_2/\sqrt 2}\cos(x_1/\sqrt 2)-x_2^2$ &
  $x_2p_1+e^{-p_2/\sqrt 2}\sin(x_1/\sqrt 2)$\\
  \hline
  $i\sin\fz$ & $p_1^2-2\cos(x_1/\sqrt 2)\sinh(p_2/\sqrt 2)-x_2^2$ &
  $x_2p_1+\sin(x_1/\sqrt 2)\cosh(p_2/\sqrt 2)$\\
  \hline \hline\end{tabular}
  \end{center}
  \centerline{
  \parbox{14cm}{
  \caption{Equivalent real Hamiltonian $h$ and the integral of motion
  $H_i$ for various ${\cal PT}$-symmetric analytic potentials $v$.
  $m$ is set to $1/2$.}
  \label{tab1}}}
  \end{table}

The invariant $H_i$ generates a set of symmetry transformations in
the phase space. The infinitesimal symmetry transformations have
the form
    \bea
    x_1&\to& x_1+ \epsilon \{x_1,H_2\}=x_1+\epsilon(\frac{x_2}{2m}),
    \label{sym1}\\
    x_2&\to& x_2+ \epsilon \{x_2,H_2\}=x_2+\epsilon~\partial_{x_1}
    v_r(
    \mbox{\footnotesize $2$}^{-\frac{1}{2}}x_1,
    \mbox{\footnotesize $2$}^{-\frac{1}{2}}p_2),
    \label{sym2}\\
    p_1&\to& p_1+ \epsilon \{p_1,H_2\}=p_1+\epsilon~\partial_{p_2}
    v_r(
    \mbox{\footnotesize $2$}^{-\frac{1}{2}}x_1,
    \mbox{\footnotesize $2$}^{-\frac{1}{2}}p_2),
    \label{sym3}\\
    p_2&\to& p_2+ \epsilon \{p_2,H_2\}=p_2-\epsilon(\frac{p_1}{2m}),
    \label{sym4}
    \eea
where $\epsilon$ is an infinitesimal real variable.

\np

\section{Summary and Conclusions}

Analytic continuation of a potential defined on the real axis to
complex plane determines a complex Hamiltonian dynamical system
having two real configurational degrees of freedom and four phase
space degrees of freedom. The condition that the symplectic
structure on the phase space $\C^2=\R^4$ be compatible with the
dynamical equations restricts the former to a four-parameter
family of symplectic structures which does not include the
standard symplectic structure. Nevertheless, all these structures
are isomorphic to the standard symplectic structure. This implies
the existence of a conventional description of the complex systems
using a real Hamiltonian that turns out to be twice the real part
of initial complex Hamiltonian $H$. The imaginary part of $H$ is
an integral of motion rendering the system integrable.

In the study of ${\cal PT}$-symmetric potentials, the imaginary
part of the classical Hamiltonian is often set to zero. This
yields certain special classical trajectories whose physical
superiority over those having $H_i\neq 0$ is not clear. The
situation resembles confining the study of the trajectories of
coulomb potential to those having a particular value of angular
momentum and ignoring the others.

It is important to note that the classical dynamics determined by
the analytic continuation of Hamilton's equations defines a
classical system whose standard canonical quantization is
different from the one corresponding to the naive prescription
    \be
    \fz\to \hat x,~~~~~~~~\fp\to\hat p,
    ~~~~~~~~~
    \{\cdot,\cdot\}\to -i[\cdot,\cdot],
    \label{canon}
    \ee
where $\hat x$ and $\hat p$ are the usual position and momentum
operators, $\{\cdot,\cdot\}$ is the Poisson bracket, and
$[\cdot,\cdot]$ is the commutator. One of the reasons for this is
that the symplectic structure associated with the Poisson bracket
is not compatible with the classical dynamical equations. As a
result the Heisenberg equations do not tend to the Hamilton
equations involving the usual Poisson bracket in the classical
limit. Another reason is that the complex classical system is
intrinsically two-dimensional (having a four-dimensional phase
space) whereas the quantum system with the Hamiltonian $\hat
H=\hat p^2/2m+v(\hat x)$ is one-dimensional. One can insist on
defining an effective one-dimensional system by enforcing $H_i=0$
as a constraint and moding out the symmetry transformations
(\ref{sym1}) -- (\ref{sym4}) it generates to construct a
two-dimensional reduced phase space
\cite{marsden-ratiu,henneaux-teitelboim}. Whether this reduced
system is related to the one corresponding to the classical limit
of the equivalent Hermitian Hamiltonian operator
\cite{jpa-2003,jpa-2004b,jpa-2005b} is an interesting question
worthy of investigation. The relation, if there is one, is
expected not to be direct, for we know that for complex analytic
potentials with a non-real spectrum there is no equivalent
Hermitian Hamiltonian operator, whereas the classical equivalent
real Hamiltonian can always be constructed.

\subsection*{Acknowledgment}

During the course of this work I have benefitted from helpful
discussions with Varga Kalantarov.

\ed